\documentclass[prd,onecolumn,notitlepage,nofootinbib,superscriptaddress,showpacs,showkeys]{revtex4-2}

\usepackage{amsfonts}
\usepackage{graphicx}
\usepackage{dcolumn}
\usepackage{amsmath}
\usepackage{amssymb}
\usepackage[T1]{fontenc}
\usepackage[utf8]{inputenc}
\usepackage{esint}
\usepackage[brazil,english]{babel}
\usepackage[usenames,dvipsnames]{color}
\usepackage{longtable}
\usepackage{ulem}
\usepackage{nicefrac}
\usepackage{bm}
\usepackage{tikz}
\usepackage{braket}
\usepackage{graphicx, nicefrac}
\usepackage{cancel}
\usepackage{mathrsfs}

\usepackage{enumitem} % Enumeracao em romanos minusculos

\begin{document}

\title{Quantum-statistical constraints on Kerr-anti-de Sitter thermodynamics}
\author{T. L. Campos}
\email{tdlimacampos@gmail.com}
\affiliation{Universidade Federal de Itajub\'{a}, Instituto de F\'{\i}sica e Qu\'{\i}mica, Aveniva BPS 1303, 37500-903, Itajub\'{a}-MG, Brazil}
    
\author{M. C. Baldiotti}
\email{baldiotti@uel.br}
\affiliation{Departamento de F\'{\i}sica, Universidade Estadual de Londrina, CEP 86051-990, Londrina-PR, Brazil}
    
\author{C. Molina}
\email{cmolina@usp.br}
\affiliation{Universidade de S\~{a}o Paulo, Escola de Artes, Ci\^{e}ncias e Humanidades, Avenida Arlindo Bettio 1000, CEP 03828-000, S\~{a}o Paulo-SP, Brazil}

\begin{abstract}
	
We develop a general framework for interpreting the thermodynamic descriptions of Kerr-anti de Sitter black holes (KadS). These descriptions satisfy a first law and respect the homogeneity required by scaling properties. Additionally, they are subject to restrictions from semiclassical arguments.
We show that temperature and angular velocity are kinematic quantities tied to a reference frame, identified through the Euclidean formalism. However, the pressure-volume contribution is a dynamical term that requires a gauge fixing of the potential mass and volume. 
It is established that the observer associated with a given thermodynamic description is directly encoded in the Killing vector that generates the horizon.
We demonstrate that the quantum statistical relation restricts the infinite family of KadS descriptions to a subclass that reduces to  Schwarzschild-adS and Kerr thermodynamics in the limits of vanishing cosmological constant and angular momentum.
Furthermore, we establish the uniqueness of both the description associated with a frame co-rotating with infinity, and the description whose thermodynamic and geometric volumes coincide.
Thus, our framework provides a coherent interpretation of the variety of KadS thermodynamics, reconciling geometric and quantum-statistical considerations.
\end{abstract}

\keywords{black-hole thermodynamics, Kerr-anti de Sitter spacetime, Euclidean formalism, quantum statistical relation, isohomogeneous transformations}
\maketitle

\section{Introduction}

The thermodynamics of black holes has long provided insights into the interplay between gravitation, quantum theory, and statistical mechanics \cite{Maldacena:1997re, Witten:1998qj, Amado:2017kao, Guica:2008mu, Brown:1994gs, Cardoso:2013pza, bardeen1973four, gauntlett1999black, townsend2001first, Gubser:1998bc}. While asymptotically flat spacetimes admit a relatively rigid thermodynamic structure, the situation is more subtle in the case of Kerr-anti de Sitter (KadS) black holes \cite{Elias:2018yct, Hemming:2007yq, Louko:1996dw, Fontana:2018drk, Baldiotti:2017ywq, chrusciel2015hamiltonian, dolan2011pressure, Lemos:2015zma}. In these spacetimes, the absence of a preferred class of stationary observers paves the way for the investigation of multiple thermodynamic descriptions, some of which have been extensively discussed in the literature.

A widely studied formulation, referred here as the usual thermodynamic theory (UTT) \cite{caldarelli2000thermodynamics}, has attracted attention for its resemblance to familiar thermodynamic systems, including Van der Waals fluids and critical phenomena \cite{gibbons2005first, dolan2012pdv, dolan2011compressibility, cvetivc2011black, xiao2024extended, kubizvnak2012p,kubizvnak2017black}, inspiring the development of black hole chemistry, where the cosmological constant plays the role of pressure, and its conjugate is a thermodynamic volume. Despite its appeal, UTT faces conceptual issues, such as mismatches between geometric and thermodynamic volumes and ambiguities in defining conserved quantities via Killing vectors \cite{cvetivc2011black,dolan2012pdv,cai2005thermodynamics}. Another possibility is Hawking's geometric approach~\cite{hawking1999rotation}, constructing thermodynamic quantities from renormalized Komar integrals. This setup, although straightforward, fails to satisfy a first law \cite{gibbons2005first}. Further efforts include extensions of the Iyer-Wald formalism~\cite{iyer1994some, gao2023general, campos2025black}, and combinations of Komar integrals with vector volumes \cite{ballik2013vector}, leading to the so-called alternative thermodynamic theory (ATT) \cite{campos2024generating}.

A unifying perspective for the different possible thermodynamic representations is offered by isohomogeneous transformations \cite{campos2024generating}, which map one thermodynamic description into another while preserving the homogeneity required by the scaling properties of the spacetime \cite{kastor2009enthalpy}. Exact isohomogeneous transformations (EITs) additionally preserve the first law, and hence generate consistent thermodynamic theories \cite{campos2025black}. In principle, this framework yields infinitely many thermodynamic descriptions of KadS black holes. The central question which motivates the present work is how to interpret this multiplicity on physical grounds. We address this issue by using arguments based on the Euclidean formalism as a central constraint to select the valid thermodynamic descriptions. Specifically, we examine the validity of the quantum statistical relation (QSR), which emerges from path-integral arguments in semiclassical field theory \cite{gibbons2005first}.

\newpage

Anticipating the main findings of the article, the following is a brief summary of our results:
\begin{enumerate}[label=\roman*)]
	
\item We show that the lack of uniqueness in the KadS thermodynamic description stems from the existence of more than two independent metric parameters, and thus thermodynamic variables. 
	
\item We demonstrate that the quantum statistical relation severely limits the viable thermodynamic descriptions, showing that they reduce to the standard Kerr thermodynamics in the limit of $\Lambda$ vanishing or to the Schwarzschild-adS thermodynamics in the static limit.
	
\item We argue that transformations between members of this thermodynamic class separate into kinematic and dynamical aspects: the former depend on the observer, while the latter also depend on a gauge choice. 

\item We show that, once a frame in KadS is specified, the associated thermodynamic description is unique, implying that the frame also fixes the gauge.

\item We establish that the UTT is uniquely selected when adopting the frame that is both co-rotating with infinity and consistent with the Hawking temperature. Conversely, the ATT is the unique description whose thermodynamic volume precisely coincides with the geometric volume, defined as the vector volume associated to the Killing generator of the ATT temperature.

\end{enumerate}

The structure of this paper is as follows. In Section~\ref{sec: therm desc} we review the KadS geometry and the main thermodynamic descriptions found in the literature. 
Section~\ref{sec:qsr} discusses the restrictions imposed by the quantum statistical relation and their consequences for KadS thermodynamics.  
Section~\ref{sec: frame} focuses on the Euclidean formalism and establishes the relation between thermodynamic descriptions and reference frames, considering temperature and black-hole angular velocity. 
Section~\ref{sec: gauge} develops the interpretation of mass and volume as gauge-dependent potential quantities, and explains how isohomogeneous transformations are implemented geometrically within this structure, subject to the QSR restriction.
Final comments are found in Section~\ref{remarks}.
In Appendix~\ref{app:functionG}, we derive the specific form of the generating functions that preserve the QSR. In Appendix~\ref{app: forms}, potential volume and other charges are reviewed. 
We use the geometric unit system and signature $(-,+,+,+)$ for the metric.

\section{Thermodynamic Descriptions for Kerr-Anti-de Sitter Black Holes}

\label{sec: therm desc}

\subsection{Kerr-anti-de Sitter spacetime}

The Kerr-anti-de Sitter (KadS) solution to the Einstein field equations describes rotating black holes within asymptotically anti-de Sitter vacuum spacetimes. These geometries are fully specified by the mass parameter~$m$, the rotation parameter~$a$, and a negative cosmological
constant~$\Lambda$, defining the parameter space~$\{(m, a, \Lambda)\}$ of
all KadS geometries.

In Boyer-Lindquist-like coordinates, the KadS metric takes the form \cite{griffiths2009exact} 
\begin{equation}
\mathrm{d}s^{2} = -\frac{\Delta _{r}}{\rho ^{2}} \left( \mathrm{d}t - \frac{
a\sin^{2}\theta \, \mathrm{d} \phi }{\Xi}\right)^{2} + \frac{\Delta_{\theta}
\sin^{2}\theta }{\rho ^{2}} \left[ a \, \mathrm{d}t - \frac{\left(
r^{2}+a^{2} \right) \, \mathrm{d}\phi }{\Xi }\right]^{2} + \frac{\rho^{2}}{
\Delta_{r}} \, \mathrm{d}r^{2} + \frac{\rho^{2}}{\Delta_{\theta }} \, 
\mathrm{d}\theta^{2} \, ,  
\label{Boyer-Lindquist Kerr-anti de Sitter}
\end{equation}
where 
\begin{align}
\begin{split}
& \Delta _{r}\equiv \frac{L^{2}+r^{2}}{L^{2}}\left( r^{2}+a^{2}\right)
-2mr~, \qquad \Delta _{\theta }\equiv 1-\frac{a^{2}}{L^{2}}\cos ^{2}\theta~,
\\
& \rho ^{2}\equiv r^{2}+a^{2}\cos ^{2}\theta ~,\qquad \Xi \equiv 1-\frac{a^{2}%
} { L^{2}}~, \qquad L^{2}\equiv -\frac{3}{\Lambda }\,.
\end{split}
\label{quantidades BL1}
\end{align}
In this work, we focus on the non-extremal regime, where the function $\Delta_r$ has two distinct positive real roots. The larger root, denoted by $r_{+}$, corresponds to the event horizon, and thus the boundary of the black hole. To ensure a regular Lorentzian signature throughout the coordinate chart, it is required that $\Xi>0$. 
The spacetime exhibits two Killing vector fields, which are, outside the event horizon $(r>r_+)$: a timelike field $\partial_{t}$ associated with stationarity, and a spacelike field $\partial_{\phi}$ corresponding to axial symmetry. 
The area of the event horizon $A$ and the mass parameter~$m$ can be written as
\begin{equation}
A = \frac{4\pi \left(r_{+}^{2} + a^{2} \right)}{\Xi}~, \qquad m = \frac{%
\left( r_+^2 + a^2 \right) \left(L^2 + r_+^2\right)}{2 r_+ L^2}~.
\end{equation}

The line element~\eqref{Boyer-Lindquist Kerr-anti de Sitter} can be
expressed in the alternative form: 
\begin{equation}
\mathrm{d}s^{2} = -\mathcal{N}^2 \, \mathrm{d}t^{2} + \frac{\rho ^{2}}{\Delta _{r}}
\, \mathrm{d}r^{2} + \frac{\rho ^{2}}{\Delta _{\theta }} \, \mathrm{d}\theta
^{2} + \frac{\Sigma^{2} \sin^{2}\theta }{\rho ^{2} \, \Xi^{2}} \left( 
\mathrm{d}\phi - \omega \, \mathrm{d}t\right) ^{2} \, ,
\label{second form KadS}
\end{equation}
which makes the angular velocity of the black hole manifest. The functions $\Sigma$, $\mathcal{N}$, and $\omega$ are defined as 
\begin{equation}
\Sigma ^{2} \equiv \left( r^{2}+a^{2} \right)^{2}\Delta _{\theta } -
a^{2}\Delta_{r}\sin^{2}\theta~, \qquad
\mathcal{N}^2  \equiv \frac{\rho ^{2}\Delta
_{r}\Delta _{\theta }}{\Sigma ^{2}} ~, \qquad 
\omega \equiv \frac{a\Xi }{\Sigma ^{2}} \left[\Delta _{\theta
}(r^{2}+a^{2})-\Delta _{r} \right]~.
\label{quantidades BL2}
\end{equation}

The quantity $\omega$ represents the angular velocity of a zero-angular-momentum observer (ZAMO). It approaches the angular velocity of the black hole at the event horizon ($\Omega_{H})$, and asymptotically approaches the angular velocity of a rotating frame at infinity~\cite{caldarelli2000thermodynamics}: 
\begin{equation}
\lim_{r\rightarrow r_{+}}\omega =\Omega _{H}~, \qquad \lim_{r\rightarrow
\infty }\omega =-\frac{a}{L^{2}}~,  
\label{w at r+ and inf}
\end{equation}
with
\begin{equation}
\Omega_{H} \equiv \frac{a\,\Xi}{a^{2}+r_{+}^{2}}~.
\end{equation}

The event horizon is a Killing horizon generated by a null Killing vector field with the form 
\begin{equation}
K^{\mu } \equiv \xi ^{\mu } + \Omega \varphi ^{\mu } \, ,  \label{killing}
\end{equation}
where $\xi^\mu$ and $\varphi^\mu$ are Killing vectors associated with time translation and axial rotation, respectively, and $\Omega$ is a constant. 
A standard choice for $\xi^{\mu}$ and $\varphi^{\mu}$ is
\begin{equation}
\xi = \partial_t ~, \qquad \varphi = \partial_\phi ~, \qquad \Omega = \Omega_H
~.  \label{coordinate basis}
\end{equation}
With this definition, the corresponding surface gravity is 
\begin{equation}
\kappa_+ = \frac{\left( L^{2}+3r_{+}^{2} \right) r_{+}^{2} - a^{2} \left(
L^{2} - r_{+}^{2} \right)} {2L^{2}r_{+} \left( r_{+}^{2} + a^{2} \right)} ~.
\label{eq:kappa}
\end{equation}

\subsection{Isohomogeneous transformations}

A striking feature of KadS black holes is their rich thermodynamic structure, which admits multiple descriptions. In this sense, KadS spacetime is fundamentally different from Kerr and Schwarzschild-adS geometries. A central result of the present work, detailed in the following sections, is the interpretation of this diversity.
Before reviewing these descriptions, we recall the concept of isohomogeneous transformations \cite{campos2024generating}, a method for generating new thermodynamic representations from an existing one.

We define a consistent thermodynamic description of KadS as a choice of thermodynamic potential $(M_0)$, temperature $(T_0)$, angular velocity $(\Omega_0)$, and volume $(V_0)$ that satisfies the first law of thermodynamics, 
\begin{equation}
\mathrm{d}M_0 = T_0 \mathrm{d}S + \Omega_0 \mathrm{d}J + V_0 \mathrm{d}P~,
\end{equation}
while sharing the common definitions for entropy, angular momentum, and
pressure:
\begin{equation}
S = \frac{A}{4}~,
\qquad
J = \frac{a m}{\Xi^2}~,
\qquad
P = \frac{3}{8\pi L^2}~.
\end{equation}
Furthermore, the thermodynamic description must be homogeneous. This implies that the variables $M_0$, $T_0$, $\Omega_0$ and $V_0$ must satisfy the Smarr relation: 
\begin{equation}
M_0 = 2 T_0S + 2 \Omega_0 J - 2 V_0 P~.  
\label{original smarr}
\end{equation}
This form of Eq.~\eqref{original smarr} is consistent with the standard scaling argument for thermodynamics \cite{kastor2009enthalpy}.

Isohomogeneous transformations are a class of maps, parameterized by two functions $g \equiv g(S,J,P)$ and \mbox{$h \equiv h (S,J,P)$} that act on an original thermodynamic description ($M_0$, $T_0$, $\Omega_0$, $V_0$) to produce a new one ($M_1$, $T_1$, $\Omega_1$, $V_1$): 
\begin{equation}
M_{1}\equiv gM_{0} ~, ~~
T_{1}\equiv gT_{0}+h \frac{\partial g}{\partial S}~, ~~
\Omega _{1}\equiv g\Omega _{0} + h \frac{\partial g}{\partial J} ~, ~~
V_{1}\equiv gV_{0}+h\frac{ \partial g}{\partial P}~.  
\label{generated thermodynamics}
\end{equation}
Homogeneity is preserved in the process, that is, $M_1$ and $M_0$ are homogeneous functions of the same degree satisfying a Smarr formula: 
\begin{equation}
M_{1} = 2T_{1}S + 2\Omega _{1}J-2V_{1}P ~.  
\label{rl}
\end{equation}
Note, however, that despite its appearance, Eq.~\eqref{rl} does not necessarily represent an Euler relation, since a first law may not hold in the new description. A counterexample demonstrates this fact in the next subsection [see Hawking's proposal, summarized in Eq.~\eqref{Hawking quantities}].

Homogeneity is ensured by choosing $g$ to be a positive definite homogeneous function of degree zero, 
\begin{equation}
S\frac{\partial g}{\partial S}+J\frac{\partial g}{\partial J}-P\frac{\partial g}{\partial P}=0~,  
\label{homoeneity of g}
\end{equation}
and $h$ to have the same homogeneity as $M_{0}$. 
The subset of exact isohomogeneous transformations (EITs), defined by $\{(g,M_{0})\}$, namely those that satisfy $h=M_0$, represents contactomorphisms in thermodynamic state space \cite{campos2025black}. In this case the first law is also preserved in the new description: 
\begin{equation}
\mathrm{d}M_{1}=T_{1}\mathrm{d}S+\Omega _{1}\mathrm{d}J+V_{1}\mathrm{d}P~.
\end{equation}
If Eq.~\eqref{rl} is an Euler relation satisfying the first law of thermodynamics, it follows that it is a thermodynamic Smarr formula~\cite{campos2024generating}. 

An equivalent representation of Eq.~\eqref{generated thermodynamics} is 
\begin{equation}
T_{1} = N T_{0}~,
\qquad 
\Omega _{1} = N \Omega _{0} - \Omega_\text{frame} ~,
\qquad 
V_{1} = N V_{0} - V_\text{gauge} ~,
\label{pqr1}
\end{equation}
with 
\begin{equation}
N=g+\frac{h}{T_{0}}\frac{\partial g}{\partial S}~,\qquad \Omega_\text{frame}=h\left( \frac{ \Omega _{0}}{T_{0}}\frac{\partial g}{\partial S}-\frac{\partial g}{\partial J}\right) ~,\qquad 
V_\text{gauge}=h\left( \frac{V_{0}}{T_{0}}\frac{\partial g}{\partial S} -
\frac{\partial g}{\partial P}\right) ~.  
\label{pqr2}
\end{equation}
As detailed in the following sections, the notation $\{ N, \Omega_{\text{frame}}, V_{\text{gauge}} \}$ in Eq.~\eqref{pqr1} reflects physical roles of the variables: $N$ is a normalization factor for the Killing field; $\Omega_\text{frame}$ is the angular velocity linked to an observer; and $V_\text{gauge}$ defines the choice of a gauge for characterizing the thermodynamic volume.

\subsection{Notable thermodynamic descriptions}
\label{td}

To fix the notation and lay the groundwork for subsequent analysis, we review three well-known thermodynamic descriptions for KadS black holes. 
It is instructive to classify these frameworks based on the integrability properties of the first law, which, more generally, assume a variational identity of the form
\begin{equation}
\delta M = \frac{\kappa}{2\pi}\delta S + \Omega \delta J + V \delta P ~.
\label{variational_law}
\end{equation}
This functional relation can be derived via extended Iyer-Wald formalisms~\cite{iyer1994some, campos2025black} and represents an inexact relation when the variations $\delta$ cannot be replaced by exterior derivatives $\mathrm{d}$.
However, for a thermodynamic description to be consistent in the standard equilibrium sense, Eq.~\eqref{variational_law} must become an exact first law formed by exact differentials.

The first description of interest, proposed by Hawking~\cite{hawking1999rotation}, provides a clear geometric foundation for the KadS thermodynamics. However, it does not constitute a complete thermodynamic theory, because it satisfies an inexact first law \cite{campos2024generating}.
Its quantities are derived from generalized Komar integrals and the vector volume formalism using the Killing generator defined in Eq.~\eqref{killing}, with the choice of the Killing fields $\xi$ and $\varphi$ in Eq.~\eqref{coordinate
basis}. 
The resulting quantities are: 
\begin{equation}
M_H = \frac{m}{\Xi}~, 
\qquad
T_H  = \frac{\kappa_+}{2\pi}~, 
\qquad
\Omega_H = \frac{a \Xi}{a^2 +r_+^2}~, 
\qquad
V_H = \frac{4\pi}{3} \frac{r_+(r_+^2+a^2)}{\Xi}~.
\label{Hawking quantities}
\end{equation}

Although Hawking's original proposal is termed inexact within the present work's classification, it can also be considered integrable, in the sense that it can be modified via an integrating factor to yield an exact first law of thermodynamics \cite{campos2024generating, gao2023general, campos2025black}. 
In this approach, the variational operators in Eq.~\eqref{variational_law} can be replaced by exact differentials. The resulting alternative thermodynamic theory (ATT) is characterized by
\begin{equation}
M_A = \frac{M_H}{\sqrt{\Xi}}~,
\qquad
T_A = \frac{T_H}{\sqrt{\Xi}}~, 
\qquad
\Omega_A = \frac{\Omega_H}{\sqrt{\Xi}}~, 
\qquad 
V_A = \frac{V_H}{\sqrt{\Xi}}~.  
\label{ATT quantities}
\end{equation}

A related and well-established consistent thermodynamic framework for KadS black holes is defined by the following set of quantities~\cite{caldarelli2000thermodynamics, gibbons2005first}:
\begin{equation}
M_U = \frac{M_H}{\Xi}~, 
\qquad 
T_U = T_H ~, 
\qquad 
\Omega_U = \Omega_H + \frac{a}{L^2}~, 
\qquad
V_U = V_H + \frac{4\pi}{3} a^2 M_U~.  
\label{UTT quantities}
\end{equation}
This representation, referred as the usual thermodynamic theory (UTT) in~\cite{campos2024generating, campos2025black}, has a less intuitive physical interpretation compared to the previous formulations. The discrepancy between the geometric and thermodynamic definitions of volume is emphasized, because it suggests a non-trivial relationship between geometric and thermodynamic structures in KadS spacetime.%
\footnote{We refer to the geometric volume of a particular thermodynamic description as that obtained through the vector volume formalism \cite{ballik2013vector} using the same Killing field that generates the surface gravity (i.e., the temperature) of that representation. In this sense, both $V_H$ and $V_A$ are geometric volumes, associated with different normalizations of the Killing vectors \cite{campos2024generating,campos2025black}.}

\begin{figure}[h]
\includegraphics{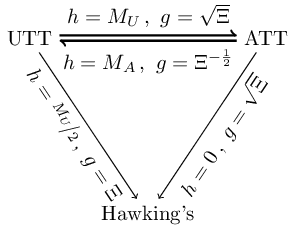}
\caption{Isohomogeneous transformations between notable descriptions for the KadS thermodynamics. The map is exact from UTT to ATT, and from ATT to UTT (thick arrows).} 
\label{fig:transformations}
\end{figure}

The relations between these thermodynamic descriptions, mediated by isohomogeneous transformations, are summarized in Fig.~\ref{fig:transformations}.

\section{Constraints from Euclidean Quantum Field Theory}
\label{sec:qsr}

\subsection{General setup}

A central result underlying black-hole thermodynamics is the quantum statistical relation (QSR) \cite{gibbons1977action}. Emerging from the formalism of Euclidean quantum field theory, the QSR links geometry and thermal physics by encoding a black hole's temperature in the structure of its spacetime. Moreover, as we have seen, isohomogeneous transformations are a powerful tool for generating new thermodynamic descriptions. In this section, we impose the validity of the QSR as a constraint on KadS thermodynamics and, consequently, on the admissible isohomogeneous transformations.

In the Euclidean approach, the KadS metric~\eqref{Boyer-Lindquist Kerr-anti de Sitter} is associated with the (vacuum) gravitational action with a cosmological constant~$\Lambda $,
\begin{equation}
I=-\frac{1}{16\pi }\int_{\mathcal{M}}\mathrm{d}^{4}x\sqrt{g}~\left(
R-2\Lambda \right) -\frac{1}{8\pi }\int_{\partial \mathcal{M}} 
\mathrm{d}^{3}x\sqrt{h}~[K]~,  
\label{einstein-maxwell action}
\end{equation}
where $R=4\Lambda$ is the Ricci scalar, and $[K]$ is the trace of the extrinsic curvature including a counterterm to renormalize the action \cite{gibbons2005first}. 
The thermal behavior of a system is described by considering a regular manifold and fields with periodic Euclidean time. 
For the KadS geometry, a regular Euclidean section\footnote{%
In stationary, non-static spacetimes, the induced metric is complex, forming
a ``quasi-Euclidean section''. For convenience, this is still commonly referred to as ``Euclidean''.} 
$\mathcal{M}$ is obtained by imposing the periodic identification~\cite{hawking1999rotation} 
\begin{equation}
\left( t~,~\phi \right) \sim \left( t+i\beta ~,~\phi +i\beta \,\, \Omega \right) ~.  
\label{Hawking ident}
\end{equation}
Acceptable values for $\beta$ are chosen by the requirement of eliminating
conical singularities. Eq.~\eqref{Hawking ident} corresponds to Hawking's description, in which the temperature is $\beta^{-1}= T_H$ and the black-hole angular velocity is $\Omega=\Omega_H$.

Let us briefly summarize the quantum statistical relation. In the saddle-point approximation, the partition function $Z$ of a thermal field is related to the Euclidean action $I$ as follows:
\begin{equation}
\ln Z = - I ~.  
\label{eq:lnZ0-1}
\end{equation}
For a grand-canonical ensemble, this is equivalently expressed through the
grand potential $W$ as 
\begin{equation}
\ln Z = -\beta W ~,
\label{eq:lnZ0-2}
\end{equation}
with $\beta \equiv T^{-1}$. In this framework, the exterior spacetime of the black hole exhibits thermal behavior \cite{gibbons1977action}, with a grand potential of the form
\begin{equation}
W = M - T S - \Omega J ~.  
\label{grand potential}
\end{equation}
Identifying the two expressions for $\ln Z$ in Eqs.~\eqref{eq:lnZ0-1}-\eqref{eq:lnZ0-2} leads to the quantum statistical relation:
\begin{equation}
M - T S - \Omega J = T I ~.  
\label{qsr}
\end{equation}
The validity of Eq.~\eqref{qsr} for the UTT was demonstrated in \cite{gibbons2005first} and for the ATT in \cite{campos2024generating}.

Using the Smarr relation \eqref{original smarr}, we can rewrite Eq.~\eqref{qsr} as
\begin{equation}
I = \beta \left( \frac{M}{2} - V P \right) ~.
\label{qsr2}
\end{equation}
In particular, for the notable thermodynamic descriptions discussed in subsection~\ref{td},
\begin{equation}
I=\beta _{A}\left( \frac{M_{A}}{2}-V_{A}P\right) =
\beta _{H}\left( \frac{M_{H}}{2}-V_{H}P\right) =
\beta _{U}\left( \frac{M_{U}}{2}-V_{U}P\right) ~.
\label{qsr-des}
\end{equation}
In Eq.~\eqref{qsr-des}, $\beta _{A} \equiv T_{A}^{-1}$, $\beta _{H} \equiv T_{H}^{-1}$, and $\beta _{U} \equiv T_{U}^{-1}$.

From the geometric point of view, the association between the ATT and the Hawking approach in Eq.~\eqref{qsr-des} follows directly from the Euclidean gravitational action~\eqref{einstein-maxwell action}. 
While the latter naturally satisfies the QSR due to its derivation from the Euclidean formalism \cite{hawking1999rotation}, the ATT relates to Hawking's description through a constant rescaling of the generating Killing vector \cite{campos2024generating}:%
\footnote{As discussed in the following section, more general transformations involve not only a rescaling of the Killing field but also a linear combination of $\partial_t$ and $\partial_\phi$.}
\begin{equation}
	K_H = \partial_t + \Omega_H \partial_\phi~, \qquad K_A = \frac{\partial_t}{\sqrt{\Xi}} + \frac{\Omega_H}{\sqrt{\Xi}}\partial_\phi~.
    \label{KH e KA}
\end{equation}

On the other hand, the last equality in Eq.~\eqref{qsr-des} does not share the same geometric nature, as $T_U$ and $T_H$ coincide, but $V_U$ and $V_H$ do not. Instead, the validity of the QSR for the UTT only follows from 
\begin{equation}
\frac{M_{U}}{2}-V_{U}P=\frac{M_{H}}{2}-V_{H}P~,
\end{equation}
which does not have a clear geometric interpretation.

While Hawking's description obeys the formal quantum statistical relation~\eqref{qsr-des}, it does not satisfy a first law and it is thus thermodynamically inconsistent. As a result, the combination $M_{H}-T_{H}S-\Omega _{H}J$ does not represent a proper thermodynamic grand potential. 
This demonstrates that the QSR is not a sufficient condition for thermodynamic consistency. Conversely, a consistent thermodynamic description may not satisfy the QSR for the gravitational action~\eqref{einstein-maxwell action}, since it may be the case that Eq.~\eqref{qsr2} does not hold. 
In the following subsection, we derive the restrictions on isohomogeneous transformations required to preserve the QSR.

\subsection{Preserving the quantum statistical relation}

Consider a thermodynamic representation (labeled by 0) that satisfies both the exact first law and the quantum statistical relation. Following Eq.~\eqref{qsr-des}, a new thermodynamic description (labeled by 1) will also satisfy the QSR if and only if 
\begin{equation}
\beta _{1}\left( \frac{M_{1}}{2}-V_{1}P\right) =I=\beta _{0}\left( \frac{%
M_{0}}{2}-V_{0}P\right) ~. \label{eq I}
\end{equation}
Substituting the isohomogeneous transformation relations from Eq.~\eqref{generated thermodynamics} into this condition leads to a constraint on the generating function 
$g$:
\begin{equation}
\frac{\partial g}{\partial S}I + \frac{\partial g}{\partial P}P=0~.
\label{condition qsr}
\end{equation}
The solutions of Eq.~\eqref{condition qsr} form the subset of isohomogeneous transformations that preserve the QSR. 

It should be noticed that, if $g$ is independent of $J$, the homogeneity condition \eqref{homoeneity of
		g} implies that 
\begin{equation}
\frac{\partial g}{\partial S} \, (I+S) = 0 ~.
\label{sads and kerr unique}
\end{equation}
Therefore $g$ must also be independent of $S$ and $P$ [using also Eq.~\eqref{condition qsr}].
This suggests that, for Schwarzschild-adS and Kerr spacetimes, there is at most one thermodynamic description satisfying the QSR. However, for KadS, the additional degree of freedom allows for other solutions.

We can go further by restricting the functional form of the functions $g$ that satisfy the QSR constraint~\eqref{condition qsr}. As shown in appendix~\ref{app:functionG}, all the functions $g$ compatible with the QSR have the functional form:
\begin{equation}
g (S,J,P) = G \left( \frac{J^{2}P}{M_{0}^{2}} \right) ~,
\label{g = g(y0)}
\end{equation}
with $G(0) = 1$. 
This last condition ensures a unique description in the non-rotating or zero-pressure limits. To better quantify this previous statement, let us define the dimensionless variable
\begin{equation}
\mathcal{F} \equiv \frac{J^{2} P}{M^{2}} ~.
\label{eq:factorF}
\end{equation}
The quantity $\mathcal{F}$ acts as a degeneracy factor for the thermodynamic representations of KadS black holes. When $\mathcal{F} = 0$ (as in Schwarzschild-adS or non-rotating cases), the thermodynamic description is unique. A non-zero $\mathcal{F}$ signals multiple consistent thermodynamic formalisms.

Adopting the notation $G' \equiv \mathrm{d} G(\mathcal{F}) / \mathrm{d} \mathcal{F}$, the
transformations for temperature, angular velocity, and volume of Eq.~\eqref{generated thermodynamics} then become 
\begin{equation}
T_{1} = NT_{0} ~,\qquad 
\Omega _{1} = N\Omega _{0} + \frac{2 h J P}{M_{0}^{2}}G' ~, \qquad 
V_{1} = NV_{0} + \frac{h J^{2}}{M_{0}^{2}} G' ~.  
\label{qsr T omega V}
\end{equation}
In the present case, the term $N$ defined in Eq.~\eqref{pqr2} reduces to
\begin{equation}
N = G - \frac{2 h J^{2} P}{M_{0}^{3}} G' ~. 
\label{qsr p}
\end{equation}

Note that $\Xi$ is a special quantity, in the sense that it has the form~\eqref{g = g(y0)},
\begin{equation}
	\Xi 
    = \Xi (\mathcal{F})
    = 1-\frac{8\pi }{3}\frac{J^{2}P}{M_{U}^{2}} =
    \left( 1+\frac{8\pi }{3} 
	\frac{J^{2}P}{M_{A}^{2}}\right) ^{-1} ~.
	\label{xi=xi(F)}
\end{equation}
Eq.~\eqref{xi=xi(F)} clarifies the role of $\Xi$: it serves as a geometric manifestation of the degeneracy factor~\eqref{eq:factorF}, appearing explicitly within the KadS metric~\eqref{Boyer-Lindquist Kerr-anti de Sitter}.

In the limit of a vanishing cosmological constant ($\Lambda \to 0$), we have $\Xi \to 1$. Consequently, the infinite family of representations parameterized by $g = G(\Xi)$ collapses to the standard thermodynamic description of a Kerr black hole. 
Similarly, in the static limit ($a \to 0$), $\Xi \to 1$ and the KadS representations reduces to the standard Schwarzschild-adS thermodynamic framework.

\section{Kinematic terms and observer dependence}
\label{sec: frame}

\subsection{Consistency of temperature and angular velocity with the first law}

The thermodynamic description of black holes relies on the identification of observer-dependent, or kinematic, variables. For the KadS spacetime, these variables are temperature and angular velocity. In this section, we will carefully examine these quantities and their relation with the various representations of KadS thermodynamics.

Indeed, the definition of temperature from the period of Euclidean time in Eq.~\eqref{Hawking ident} is coordinate-dependent, since different time coordinates have different periods. This is not surprising, given the well-known case of static observers in Schwarzschild spacetime, where a time transformation can be used so that the periodicity agrees with the Tolman redshift factor~\cite{gibbons1978black, santiago2018tolman}. 

However, we will demonstrate that the permitted periodicities within the Euclidean formalism are not arbitrary. These periodicities are constrained to produce temperatures and angular velocities that are consistent with a thermodynamic description, linking local horizon quantities to those measured by a particular observer.
For instance, consider the rescaled time coordinate 
\begin{equation}
	t_{A}\equiv t\sqrt{\Xi }~.
\end{equation}
The proper identification for a regular Euclidean section~\eqref{Hawking ident} becomes 
\begin{equation}
	(t_{A}~,~\phi )\sim (t_{A}+i\beta _{A}~,~\phi +i\beta _{A}\Omega _{A})~,
\end{equation}
with $\beta _{A} \equiv T_{A}^{-1}$.
The quantities $T_{A}$ and $\Omega_{A}$ are given in Eq.~\eqref{ATT quantities}, associated with the ATT.

In addition to rescaling time, one can also consider different rotating
frames. Starting from the ATT, let us define new coordinates 
\begin{equation}
	t_1 \equiv N^{-1} \, t_A ~, \qquad \phi_1 \equiv \phi - \Omega_\text{frame} \, t_1~.
	\label{p,q}
\end{equation}
Here, $N$ and $\Omega_\text{frame}$ are spacetime scalars, which we will shortly connect to
Eqs.~\eqref{pqr1}-\eqref{pqr2}. The new identification for the Euclidean
section is 
\begin{equation}
	(t_1~,~\phi_1) \sim (t_1 + i \beta_1~,~\phi_1 + i \beta_1 \Omega_1)~,
	\label{new ident}
\end{equation}
with 
\begin{equation}
	\beta_1 \equiv N^{-1}\, \beta_A ~, \qquad \Omega_1 \equiv N \Omega_A - \Omega_\text{frame}~.
	\label{new beta1 Omega1}
\end{equation}

As discussed in Section~\ref{sec: therm desc}, new thermodynamic descriptions can be generated from the ATT via isohomogeneous transformations. From the Euclidean formalism, the temperature of the new description is associated with the time coordinate 
\begin{equation}
	t_{1}=t_{A}\left( G + \frac{ h }{T_{A}}\frac{\partial \, G}{\partial S}\right)^{-1}~.  
    \label{tA to t1}
\end{equation}
This ensures that the periodicity is $t_{1}\sim t_{1}+i\beta _{1}$, with the appropriate temperature of Eq.~\eqref{pqr2},
\begin{equation}
	\beta _{1}=\beta _{A}\left( G+\frac{h}{T_{A}}\frac{\partial \, G}{\partial S%
	}\right) ^{-1}~.
\end{equation}
Similarly, the black hole's angular velocity is measured with respect to the
rotating angular coordinate,
\begin{equation}
	\phi _{1}=\phi - h\left( \frac{\Omega _{A}}{T_{A}}\frac{\partial \, G}{\partial S}-\frac{\partial \, G}{\partial J}\right) t_{1}~,
	\label{phi1 to phiA}
\end{equation}
which ensures that the periodicity is $\phi _{1}\sim \phi _{1}+i\beta
_{1}\Omega _{1}$, with 
\begin{equation}
	\Omega _{1}=\left( G+\frac{h}{T_{A}}\frac{\partial \, G}{\partial S}\right)
	\Omega _{A}-h\left( \frac{\Omega _{A}}{T_{A}}\frac{\partial \, G}{\partial S}-\frac{\partial \, G}{\partial J}\right) =\, G\Omega _{A}+h\frac{\partial G}{\partial J}~.
\end{equation}
As anticipated, results~\eqref{tA to t1} and~\eqref{phi1 to phiA} are the constraints of the Euclidean periodicities induced by the thermodynamics. 
For a generic choice of $h$ with the same degree of homogeneity as $M_A$, the resulting description satisfies the QSR and has a $M_1$ that inherits this homogeneity. In the particular case of $h = M_A$, this is an exact isohomogeneous transformation, thus $M_1$ becomes a well-defined thermodynamic potential, satisfying an exact first law.

Let us consider general isohomogeneous transformations, starting from an initial Euclidean identification
\begin{equation}
	(t_{0}~,~\phi _{0})\sim (t_{0}+i\beta _{0}~,~\phi _{0}+i\beta _{0}\Omega
	_{0})~.
    \label{eq:initial-description}
\end{equation}
In Eq.~\eqref{eq:initial-description}, $\{t_{0},\phi _{0}\}$ are the coordinates adapted to the original thermodynamic description, not constants. Under the coordinate transformation 
\begin{equation}
	t_{1}=t_{0}N^{-1}~, \qquad \phi _{1}=\phi _{0}-\Omega_\text{frame}t_{1}~,
	\label{t0 and phi0 to t1 and phi1}
\end{equation}
with the parameters $N$ and $\Omega_\text{frame}$ defined as in Eq.~\eqref{pqr2}, the periodicity transforms to
\begin{equation}
	(t_{1}~,~\phi _{1})\sim (t_{1}+i\beta _{1}~,~\phi _{1}+i\beta _{1}\Omega
	_{1})~,  \label{new identification}
\end{equation}
where the new inverse temperature $\beta_{1}$ and black-hole angular velocity $\Omega_{1}$ coincide with those given in Eq.~\eqref{generated thermodynamics} with Eq.~\eqref{g = g(y0)}. 
This is a new result in the section: isohomogeneous transformations are implemented in the Euclidean formalism through a corresponding change of periodic coordinates.

\subsection{Reference frames and observers}

We now interpret previous results in terms of reference frames linked to physical observers. Consider the time and angular coordinates $\{t_{0} , \phi_{0} \}$ associated with a selected reference frame. Again, the subscript 0 indicates that these coordinates are adapted to an original thermodynamic description. We define the curve 
\begin{equation}
	\alpha: \, t_1 \longmapsto 
    \big( N\, t_1\, , \, r \, , \, \theta \, , \, \Omega_\text{frame}\, t_1 \big)~, \qquad 
    \text{$r$ and $\theta$ fixed ~.}  
    \label{curve}
\end{equation}
The parameter of the curve $\alpha$ is set as $t_1$, introduced in Eq.~\eqref{t0 and phi0 to t1 and phi1}, and is chosen such that $\phi_1$ is a constant (set to zero for convenience). Thus, the curve is static in the new frame $\{t_1,\phi_1\}$, with $t_1 = t_1(t_0)$ and $\phi_1 = \phi_1(t_0,\phi_0)$. The non-null four-velocity components of $\alpha$ are
\begin{equation}
	\frac{\mathrm{d}t_0}{\mathrm{d}t_1} = N =  
    G + \frac{h}{T_0} \frac{\partial
	\, G}{\partial S}~, 
        \qquad
\frac{\mathrm{d}\phi_0}{\mathrm{d}t_1} = \Omega_\text{frame} =
	h \left( \frac{\Omega_0}{T_0} \frac{\partial \, G}{\partial S} 
    - \frac{\partial \, G}{\partial J} \right) ~.  
        \label{ang vel curve}
\end{equation}
The expression for $\mathrm{d}\phi_0/\mathrm{d}t_1$ in Eq.~\eqref{ang vel curve} can be interpreted as the angular velocity of the curve with respect to the~$\{ t_1,\phi_0 \}$ coordinates.

The angular velocity $\Omega_1$ of the black hole relative to the new frame can be calculated by first finding the angular velocity of a ZAMO with respect to the $(t_1,\phi_0)$ coordinates, evaluating it at the horizon, and then subtracting the angular velocity of the frame itself. This procedure
yields 
\begin{equation}
	\Omega_1 = \bigg( G + \frac{h}{T_0} \frac{\partial \, G}{\partial S} \bigg) %
	\Omega_0 - h \bigg( \frac{\Omega_0}{T_0} \frac{\partial \, G}{\partial S} - 
	\frac{\partial \, G}{\partial J} \bigg) = G \Omega_0 + h \frac{\partial \, G}{	\partial J}~.  
    \label{omega1 found}
\end{equation}
Given the form of $G$ in Eq.~\eqref{g = g(y0)}, $\Omega_1$ coincides with the angular velocity of the new thermodynamic description~\eqref{generated thermodynamics}.
It is straightforward to check that the same result can be found computing $\omega_1$ in the new coordinates of Eq.~\eqref{t0 and phi0 to t1 and phi1}, 
\begin{equation}
	- \frac{g_{t_1 \phi_1}}{g_{\phi_1 \phi_1}} = - \frac{\partial t_0}{\partial
		t_1} \frac{g_{t_0 \phi_0}}{g_{\phi_1 \phi_1}} - \frac{\partial \phi_0}{\phi
		t_1} \frac{g_{t_0 \phi_0}}{g_{\phi_1 \phi_1}}~,
\end{equation}
and evaluating it at the horizon.

We conclude that the thermodynamic description $(T_0, \Omega_0, V_0)$ is linked to observers associated with coordinates $(t_0, r, \theta, \phi_0)$, for which the Euclidean section has periods $i\beta_0$ and $i\beta_0 \Omega_0$. A different description $(T_1, \Omega_1, V_1)$ corresponds to observers with the trajectory given by Eq.~\eqref{curve}, which is conveniently associated with coordinates $(t_1, r, \theta, \phi_1)$.

The new reference frame is encoded in the Killing field that generates the thermodynamic description. In adapted coordinates, consider that the original representation has temperature $T_{0}$ given by the surface gravity generated by 
\begin{equation}
	K_{0}\equiv \partial _{t_{0}}+\Omega _{0}\partial _{\phi _{0}}~.  \label{K0}
\end{equation}
The induced description with temperature $T_{1}$ is then associated with the rescaled Killing field $K_{1}=N K_{0}$. Using the definitions of $N$ and $\Omega_\text{frame}$, $K_{1}$ can be rewritten as 
\begin{equation}
	K_{1}=\underbrace{\left[ \left( G+\frac{h}{T_{0}}\frac{\partial \, G}{\partial S%
		}\right) \partial _{t_{0}}+h\left( \frac{\Omega _{0}}{T_{0}}\frac{\partial \, G%
		}{\partial S}-\frac{\partial \, G}{\partial J}\right) \partial _{\phi _{0}}%
		\right] }_{\partial _{t_{1}}}+\underbrace{\left( G\Omega _{0}+h\frac{%
			\partial \, G}{\partial J}\right) }_{\Omega _{1}}\partial _{\phi _{0}}~.
	\label{K1}
\end{equation}
The first term $(\partial _{t_{1}})$ on the right-hand side of Eq.~\eqref{K1} is the generator of the curves defined by Eq.~\eqref{curve}, adapted to the $\{t_1, \phi_1\}$ frame. From this construction, an observer for whom the black hole has temperature $T_1$ and spins with angular velocity $\Omega_1$ follows a trajectory given by $\alpha$ in  Eq.~\eqref{curve}.
Furthermore, for exact transformations ($h=M_0$), $K_1$ agrees with \cite{campos2025black} and $\partial _{t_{1}}$ defines the frame of a proper thermodynamic representation, i.e., one that satisfies an exact first law for KadS black holes. This is a central result of this section.

\subsection{The usual KadS thermodynamics} 
\label{subsec: usual KadS frame}

Let us take the ATT as the original description. Its associated Killing field is \cite{campos2024generating} 
\begin{equation}
	K_{0}=K_{A}=\frac{\partial _{t}}{\sqrt{\Xi }}+\frac{%
		\Omega _{H}}{\sqrt{\Xi }}\partial _{\phi }~.
\end{equation}
As shown in Figure~\ref{fig:transformations}, the EIT from ATT to UTT is characterized by 
\begin{equation}
	\left( g,h\right) =\left( \frac{1}{\sqrt{\Xi }},M_{A}\right) ~.
	\label{ATT to UTT}
\end{equation}
From Eq.~\eqref{K1}, the Killing field that generates the UTT is
conveniently expanded in the form 
\begin{equation}
	K_{U}=\left( \partial _{t}-\frac{a}{L^{2}}\partial _{\phi }\right) +\left( 
	\frac{\Omega _{A}}{\sqrt{\Xi }}+\frac{ar_{+}^{2}}{L^{2}(a^{2}+r_{+}^{2})}%
	\right) \partial _{\phi }~,
\end{equation}
which can be rewritten as 
\begin{equation}
	K_{U}=\left( \partial _{t}-\frac{a}{L^{2}}\partial _{\phi }\right) +\left(
	\Omega _{H}+\frac{a}{L^{2}}\right) \partial _{\phi }~.
	\label{KU}
\end{equation}
The previous expression is exactly the Killing field related to Hawking's approach in Eq.~\eqref{KH e KA}, but expressed with a different expansion. The preceding analysis of reference frames supports the notion that the UTT is the thermodynamic description for KadS that is relative to the trajectories
\begin{equation}
	\alpha :\,t\longmapsto \left( t\,,\,r\,,\,\theta \,,\,-\frac{a}{L^{2}} t\right)
    ~, \qquad \text{$r$ and $\theta$ fixed}
    ~.
    \label{eq:curve-alpha}
\end{equation}
Due to Eq.~\eqref{w at r+ and inf}, the idea that this thermodynamic description corresponds to a co-rotating frame with infinity \cite{caldarelli2000thermodynamics} is formalized.

Following the development of the previous subsection, the curve $\alpha$ in Eq.~\eqref{eq:curve-alpha} is naturally adapted to the frame
\begin{equation}
	\left\{ t~,~ \phi + \frac{a}{L^2} t \right\} ~, 
	\label{UTT frame}
\end{equation}	
with $\{t,\phi\}$ representing the frame related to Hawking's approach and used to represent the KadS metric in Eq.~\eqref{Boyer-Lindquist Kerr-anti de Sitter}. Within the Euclidean formalism, the fact that both descriptions are associated with curves (observers) parameterized by the same coordinate time $t$ explains why their temperatures  coincide (i.e., $T_H = T_U$).

The quantum statistical relation is a fundamental part of the Euclidean formalism. However, the direct consequences for the connection between the KadS thermodynamic descriptions and different frames have not been fully exploited in this section.
We will now demonstrate that it is impossible to construct multiple consistent descriptions within a single reference frame.

As shown in the previous subsection, a change of reference frame is determined by the parameters $N$ and $\Omega_\text{frame}$, which are restricted by Eqs.~\eqref{qsr T omega V}-\eqref{qsr p}. First, we consider a transformation that preserves the timescale, which means setting~$N=1$. From Eq.~\eqref{qsr p}, this condition fixes $h$ as
\begin{equation}
	h = \frac{M_{0}^{3}}{2 J^{2} P} 
	\frac{G - 1}{G'}
	~.
	\label{h qsr}
\end{equation}
Substituting into Eq.~\eqref{qsr T omega V}, we obtain
\begin{equation}
T_{1}=T_{0}~, \qquad 
\Omega_{1}=\Omega _{0} + \Omega_\text{frame} ~, \qquad
\Omega_\text{frame} = \frac{M_0}{J}(G-1)~.
\label{qsr T omega V 2}
\end{equation}
To ensure that the transformation preserves the frame, we must also set $\Omega_\text{frame}=0$, resulting in $G=1$. In this case, the thermodynamic description remains unchanged. In particular, it follows that the usual KadS thermodynamics (UTT) is unique in the frame co-rotating with infinity [Eq.~\eqref{UTT frame}].

However, uniqueness in a particular frame does not preclude the existence of different thermodynamic descriptions with the same temperature. For instance, as with the UTT, other approaches also reproduce the Hawking temperature $T_H$. Indeed, even when the transformation is exact, i.e. when the first law remains exact, other representations exist for $N=1$. In this case, in which $h=M_0$ in Eq.~\eqref{h qsr}, the generating function takes the form
\begin{equation}
G = 1 + A \frac{J\sqrt{P}}{M_{0}}~,
\label{g qsr}
\end{equation}
and the new frame is that of Eq.~\eqref{qsr T omega V 2} with
\begin{equation}
\Omega_\text{frame} = A \sqrt{P}~,
\end{equation}
where $A$ is an arbitrary constant.

We conclude this section by summarizing its main results. First, the periodicities from the Euclidean formalism can be constrained to ensure consistency with homogeneity, and the first law of thermodynamics. 
Second, the terms $TS$ and $\Omega J$ in the first law represent kinematic contributions, and are fully determined by the Killing vector, which encodes the reference frame of the thermodynamic description.
Third and finally, 
the QSR fixes the correspondence between the reference frame and the KadS thermodynamic description. Consequently, the UTT is the unique representation satisfying the QSR in the frame co-rotating with infinity.

\section{Dynamical terms and gauge dependence}

\label{sec: gauge}

\subsection{Potential volume and potential mass}

We have shown that the kinematic variables, temperature and angular velocity, uniquely determine an observer and its associated frame. However, the volume and energy remain undetermined. Indeed, the thermodynamic definitions of a black hole's volume and energy have been formulated in various ways in the literature. Our work adopts the approach outlined in~\cite{campos2025black, kubizvnak2017black}, in which the definitions of mass and volume are inherently ambiguous, depending on a gauge choice. In this section, we will explore this gauge freedom and the role of the QSR.

We begin with the Smarr formula, which can be derived by integrating the following relation between a Killing vector $(\chi^\mu)$ and the Ricci tensor $(R^{\mu\nu})$: 
\begin{equation}
\nabla_\mu \nabla^\nu \chi^\mu = R^{\mu\nu} \chi_\mu~.  \label{der smarr}
\end{equation}
Let $\Sigma$ be a hypersurface extending
from $r=0$ to $r = r_0$, bounded by a 2-surface $A$ at $r = r_0$. From Eq.~\eqref{der smarr}, the quantity
\begin{equation}
-\frac{1}{2}\int_A \nabla_\mu \chi_\nu \mathrm{d}A^{\mu\nu} +\Lambda
\int_\Sigma \chi_\nu \mathrm{d}\Sigma^\nu  
\label{quant gauge independ}
\end{equation}
is independent of $r_0$.%
\footnote{For vacuum spacetimes with vanishing cosmological constant, Stokes' theorem implies that the integral of $\nabla_\nu \chi_\mu$ is independent of the 2-surface of integration, allowing one to relate quantities at infinity to those at the horizon. Eq.~\eqref{quant gauge independ} generalizes this result.}
For a timelike Killing vector $\xi^\mu$, the second term defines a vector volume \cite{ballik2013vector}, 
\begin{equation}
V_\text{vec} = \int_\Sigma \xi_\mu \mathrm{d}\Sigma^\mu~,
\end{equation}
and Eq.~\eqref{quant gauge independ} defines a generalized Komar mass \cite{campos2024generating}: 
\begin{equation}
M_\text{vec} \equiv -\frac{1}{8\pi }\int_A \nabla _{\mu }\xi _{\nu }\mathrm{%
d}A^{\mu \nu }+\frac{\Lambda }{4\pi }\int_\Sigma \xi_\nu \mathrm{d}\Sigma^{\nu }~.
\end{equation}

Alternatively, since the right-hand side of Eq.~\eqref{der smarr} is a conserved current, as it is becomes proportional to $g^{\mu \nu }\chi _{\mu }$, it can be written locally as the divergence of a 2-form. This follows from Poincar\'{e}'s lemma. In vacuum, the 2-form~$\omega _{\mu \nu }$ is a Killing potential, 
$\chi ^{\nu } = \nabla _{\mu }\omega ^{\mu \nu }.$
For KadS, Eq.~\eqref{der smarr} then implies that 
\begin{equation}
-\frac{1}{2}\int_{A} 
\left(\nabla _{\mu }\chi _{\nu }+\Lambda \omega _{\mu \nu
} \right) \,
\mathrm{d}A^{\mu \nu }  
\label{quant gauge dep}
\end{equation}
is independent of the choice of 2-surface $A$. For a timelike Killing field, the second term defines a so-called potential volume \cite{kubizvnak2017black}. When $A$ is a spatial section of the horizon, this becomes the potential volume of the black hole:
\begin{equation}
V_{\text{pot}}\ =-\frac{1}{2}\int_{A}\omega _{\mu \nu } \, \mathrm{d}A^{\mu \nu
}~.
\end{equation}

At this point, the gauge freedom of the theory becomes apparent. Since $A$ is only part of the boundary of $\Sigma$, $V_{\text{pot}}$ generally differs from $V_{\text{vec}}$. Furthermore, the quantity $\omega ^{\mu \nu }$ is not unique, that is, any divergence-free 2-form $\lambda _{\mu \nu }$ can be added to it: 
\begin{equation}
\xi ^{\nu }=\nabla _{\mu }\big(\omega ^{\mu \nu }+\lambda ^{\mu \nu }\big)~.
\label{gauge freedom in pot volume}
\end{equation}
This gauge freedom means that multiple potential volumes can be associated with the same vector volume. Thus, a different generalized Komar mass for KadS can be defined using the potential volume \cite{kastor2009enthalpy}: 
\begin{equation}
M_{\text{pot}}\ =-\frac{1}{8\pi }\int_{A}\big(\nabla _{\mu }\xi _{\nu}+\Lambda \omega _{\mu \nu }\big)\mathrm{d}A^{\mu \nu }~. \label{def Mpot}
\end{equation}
Therefore, we have 
\begin{equation}
V_{\text{vec}}\ \neq V_{\text{pot}}~,~M_{\text{vec}}\ \neq M_{\text{pot}%
}~.
\end{equation}%
Both $V_{\text{pot}}$ and $M_{\text{pot}}$ are ambiguous until a gauge is fixed. 

One possible way to fix the gauge is to require that the potential mass $M_{\text{pot}}$ coincides with a physical mass defined by other methods \cite{kubizvnak2017black}. Hereafter, we will work exclusively with potential mass and volume, denoted by $M$ and $V$, respectively. Given the developments in this article, they will be indexed by $0$ or $1$.
Appendix~\ref{app: forms} presents a more general and formal discussion of the definition of conserved charges from divergent-free vector fields using differential forms.

\subsection{Consistency of the gauge choice with the first law}

To incorporate the notions of potential mass and volume into a thermodynamic framework, we consider a Killing field $K_0$ that generates the exact first law of black-hole thermodynamics. This field is a linear combination of the Killing vectors for stationarity ($\xi _{0}$) and axial symmetry ($\varphi _{0}$),
\begin{equation}
K_{0}^{\mu }=\xi _{0}^{\mu }+\Omega _{0}\varphi _{0}^{\mu }~.  \label{K0=}
\end{equation}
In Eq.~\eqref{K0=}, the subscript $0$ on $K_{0},\xi _{0}$, and $\varphi _{0}$ labels the original thermodynamic description and is not a tensor index (the same applies to $K_{1},\xi _{1}$, and $\varphi _{1}$ in the following).
The Smarr formula is written as 
\begin{equation}
M_{0}-2\Omega _{0}J=\underbrace{-\frac{1}{8\pi }\int_{A}\nabla ^{\mu
}K_{0}^{\nu }\mathrm{d}A_{\mu \nu }}_{2T_{0}S}-\underbrace{\frac{\Lambda }{8\pi }\int_{A}\omega _{0}^{\mu \nu }\mathrm{d}A_{\mu \nu }}_{2V_{0}P}~,
\end{equation}
with 
\begin{equation}
M_{0}=-\frac{1}{8\pi }\int_{A}(\nabla ^{\mu }\xi _{0}^{\nu }+\Lambda \omega_{0}^{\mu \nu })\mathrm{d}A_{\mu \nu }~,
\quad
V_{0}=-\frac{1}{2} \int_{A}\omega _{0}^{\mu \nu }\mathrm{d}A_{\mu \nu }~.  
\label{def M_0 V_0}
\end{equation}

Note that only the Killing potential $\omega _{0}^{\mu \nu }$ of the vector field $\xi_0 ^{\mu }$ appears in the Smarr formula because $\varphi _{0}^{\mu }$ is orthogonal to hypersurfaces $\Sigma$ of constant time. This implies that 
\begin{equation}
J =\frac{1}{16\pi} 
\int_{A}\nabla ^{\mu }\varphi _{0}^{\nu }\mathrm{d}A_{\mu \nu}~.  
\label{J}
\end{equation}
The freedom to choose the gauge of the potential volume allows us to match the energy to a desired thermodynamic potential \cite{kubizvnak2017black}. Therefore, we can ensure that a new energy $M_{1}$ is related to $M_{0}$ via an EIT. As we now demonstrate, specific gauge choices lead precisely to the isohomogeneous transformations defined by Eq.~\eqref{generated thermodynamics}. 

In a previous work \cite{campos2025black}, we used an extended Iyer-Wald formalism to show how a new thermodynamic description can be generated from an initial one associated with the Killing field $K_0$. This transformation is implemented by a new Killing field, $K_1$, which takes the form of Eq.~\eqref{K1}, after the QSR restriction.
We now analyze the consequences of this transformation, specifically how $K_{1}$ affects the Komar integrals and the Smarr formula. To simplify the notation, let us define: 
\begin{equation}
\xi _{1}^{\mu }\equiv N\,\xi _{0}^{\mu }+\Omega_\text{frame}\,\varphi _{0}^{\mu }~,\quad 
~~
\varphi _{1}^{\mu }\equiv \varphi _{0}^{\mu }~,
\end{equation}
where $N$ and $\Omega_\text{frame}$ are given by Eq.~\eqref{pqr2}, with $G$ replacing $g$. Then, using $\Omega _{1}$
from Eq.~\eqref{generated thermodynamics}, Eq.~\eqref{K1} takes the form 
\begin{equation}
K_{1}^{\mu }=\xi _{1}^{\mu }+\Omega _{1}\varphi _{1}^{\mu }~.
\end{equation}

The transformation 
\begin{equation}
K_{0}^{\mu }\longrightarrow N\,K_{0}^{\mu }=K_{1}^{\mu }
\label{transforming K}
\end{equation}
implies that we must also rescale the potential $\omega _{0}^{\mu \nu }$ by the same factor $N$. However, gauge freedom allows for an additional change 
\begin{equation}
\omega _{0}^{\mu \nu }
\longrightarrow 
\omega _{1}^{\mu \nu } \equiv
N\,\omega _{0}^{\mu \nu }+\lambda_{1}^{\mu \nu } ~,  
\label{gauge transformation}
\end{equation}
with any gauge term $\lambda _{1}^{\mu \nu}$ satisfying $\nabla _{\mu }\lambda _{1}^{\mu \nu }=0$. The isohomogeneous
transformation is thus implemented geometrically by defining the new energy and volume as 
\begin{equation}
M_{1}=-\frac{1}{8\pi }\int_{A}(\nabla ^{\mu }\xi _{1}^{\nu }+\Lambda \omega_{1}^{\mu \nu })\mathrm{d}A_{\mu \nu }~,
~~
V_{1}=-\frac{1}{2} \int_{A}\omega _{1}^{\mu \nu }\mathrm{d}A_{\mu \nu }~,
\end{equation}
where $\lambda _{1}^{\mu \nu }$ is fixed by the requirement that $M_{1}=G M_{0}$, as required by Eq.~\eqref{generated thermodynamics}.

To determine the explicit form of the gauge term $\lambda _{1}^{\mu \nu }$, we evaluate the integral of the mass term $M_{1}$. Using the original Smarr formula~\eqref{original smarr} and the homogeneity of $G$ in Eq.~\eqref{homoeneity of g}, we find 
\begin{equation}
M_{1}=G M_{0}+2h\left( \frac{\partial G}{\partial P}-\frac{V_{0}}{T_{0}}\frac{\partial G}{\partial S}\right) P+P\int_{A}\lambda _{1}^{\mu \nu }\mathrm{d} A_{\mu \nu }~.  
\label{calc1}
\end{equation}
Requiring that $M_{1} = G M_{0}$ fixes $\lambda _{1}^{\mu \nu }$ such that
\begin{equation}
\lambda _{1}^{\mu \nu }=-\frac{V_\text{gauge}}{4\pi M_{0}}\left( \nabla ^{\mu }\xi
_{0}^{\nu }+\Lambda \omega _{0}^{\mu \nu }\right) ~,  \label{fixed gauge-1}
\end{equation}
revealing that the quantity $V_\text{gauge}$ from Eq.~\eqref{pqr2} is determined by this gauge-fixing procedure. 
Given an original thermodynamic description of potential $M_0$, associated with $\xi^\nu_0$ and $\omega_0^{\mu\nu}$, Eq.~\eqref{fixed gauge-1} constrains the allowed gauge forms so as to preserve the original homogeneity once a transformation is selected. If the applied isohomogeneous transformation is exact ($h = M_{0}$), then
\begin{equation}
\lambda _{1}^{\mu \nu }=\frac{1}{4\pi }\left( \frac{V_{0}}{T_{0}}\frac{%
\partial G}{\partial S}-\frac{\partial G}{\partial P}\right) \left( \nabla^{\mu }\xi _{0}^{\nu }+\Lambda \omega _{0}^{\mu \nu }\right) ~.
\label{fixed gauge}
\end{equation}

As a consistency check, we calculate the new black-hole volume: 
\begin{equation}
V_{1} = -\frac{1}{2}\int_{A}\omega _{1}^{\mu \nu }\mathrm{d}A_{\mu \nu
}=N\,V_{0}-V_\text{gauge}=GV_{0}+h\frac{\partial G}{\partial P}~.
\label{bh_volume}
\end{equation}
Result~\eqref{bh_volume} coincides with the result of the isohomogeneous transformations in Eq.~\eqref{generated thermodynamics}, subject to the QSR restriction. It can also be verified that the new Smarr formula is written as
\begin{equation}
M_{1}-2\Omega _{1}J=\underbrace{-\frac{1}{8\pi }\int_{A}\nabla ^{\mu
}K_{1}^{\nu }\mathrm{d}A_{\mu \nu }}_{2T_{1}S}-\underbrace{\frac{\Lambda }{8\pi }\int_{A}\omega _{1}^{\mu \nu }\mathrm{d}A_{\mu \nu }}_{2V_{1}P}~.
\label{def M_1 V_1}
\end{equation}

Concerning the EIT, the analysis in this section leads to interesting relations between geometry and thermodynamics. From Eq.~\eqref{transforming K}, it is straightforward to see that a transformed Killing field with the same orientation as $K_{0}^{\mu }$ outwards the black hole implies that 
\begin{equation}
T_{1 }= N T_0 > 0 ~.
\label{T1-EIT}
\end{equation}
For the result~\eqref{T1-EIT}, we are assuming that the original description has positive temperature. 
Also, from Eq.~\eqref{fixed gauge}, in order for the gauge to be generated by a future-directed timelike Killing field, as it is assumed for $\xi_{0}^{\mu }$, we must have 
\begin{equation}
\frac{1}{V_{0}}\frac{\partial G}{\partial P}>\frac{1}{T_{0}}\frac{\partial G}{\partial S}~.
\end{equation}
Due to the positivity of $T_0$ and $T_{1}$,
\begin{equation}
\frac{1}{V_{0}}\frac{\partial G}{\partial P}>-\frac{G}{M_{0}}~,
\end{equation}
and hence, assuming that the volume is positive definite in the original description, we conclude that
\begin{equation}
V_{1}=GV_{0}+M_{0}\frac{\partial G}{\partial P}>0~.
\end{equation}
Summarizing, we have found that EITs connect thermodynamically consistent theories (positive-definite temperature and volume) if they are geometrically implemented considering future-directed vectors.

\subsection{The alternative KadS thermodynamics}

In this section, we have demonstrated that the functional form of the dynamical terms depends not only on the choice of Killing vector but also on a specific gauge choice. Assuming the first law is valid within a new description, the gauge is uniquely determined by Eq.~\eqref{fixed gauge}.

This framework provides a clear interpretation of the well-known discrepancy between thermodynamic and geometric volumes in the UTT \cite{dolan2011pressure, gibbons2005first, campos2024generating, campos2025black, xiao2024extended}. 
In this formalism, a thermodynamic representation associated with a Killing field $K_0$ admits a geometric interpretation of volume: the ``vector volume'' generated by $K_0$ \cite{ballik2013vector}. 
If this geometric volume coincides with $V_0$, then any rescaling $NV_0$ remains a geometric volume for the generating Killing field $K_1 = N K_0$ of Eq.~\eqref{transforming K}. 
On the other hand, the thermodynamic volume $V_1$ may include an additional gauge term $V_\text{gauge}$. To illustrate, take $V_A$ as our geometric reference $V_0$, noting that the ATT is the unique description naturally associated with the frame
\begin{equation}
\left\{\frac{t}{\sqrt{\Xi}}~,~ \phi\right\} ~.
\end{equation}
It follows that the UTT parameters $N$ and $V_\text{gauge}$ are fixed by the transformation shown in Figure~\ref{fig:transformations}. Here, $N V_A$ coincides with the vector volume $V_H$. This occurs because the UTT and Hawking's approach use the Killing generator with the same normalization [Eqs.~\eqref{KH e KA} and~\eqref{KU}], yet the thermodynamic volume $V_U$ includes the gauge shift $V_\text{gauge}$ [Eq.~\eqref{UTT quantities}].
This observation raises the question of whether there exist other thermodynamic descriptions in which the thermodynamic volume coincides with the geometric (vector) volume. Since the ATT is a representative case of this type of correspondence, we take it as a starting point to search for others. 

The desired description would possess a temperature $N T_A$, with $N$ given by Eq.~\eqref{qsr p}, being generated by the Killing vector $N K_A$. 
Consequently, the associated geometric volume would simply be the rescaling $N V_A$. However, according to Eq.~\eqref{qsr T omega V}, this equality is only possible if $h = 0$ or $G' = 0$. 
In the former case, the transformation is ``non-exact'', leading to an inexact first law (which includes Hawking's original proposal). 
For $G' = 0$, exact cases would be possible if $h = M_A$. However, 
as demonstrated in Appendix~\ref{app:functionG}, every non-trivial alternative description necessarily follows the form of $G$ given by Eq.~\eqref{g = g(y0)}. 

This leads to a significant result: the ATT is the unique description with a thermodynamic volume identical to the black hole's geometric volume that maintains the homogeneity of the scaling argument, satisfies an exact first law, and fulfills the quantum statistical relation.

\section{Final Remarks}
\label{remarks}

In this work we have developed a general and unified framework for interpreting the different thermodynamic descriptions of Kerr-anti de Sitter black holes. Our approach follows from combining the Euclidean formalism with isohomogeneous transformations, which determine homogeneous thermodynamic representations consistent with scaling arguments. As a result, a clear separation between the kinematic and dynamical contributions of the thermodynamic quantities emerges. 

We argue that the Euclidean formalism highlights the kinematic nature of temperature and angular velocity, both of which are fully determined by the choice of reference frame. 
In addition, thermodynamic consistency requires restricting the allowed periodicities of the Euclidean time and the angular coordinate corresponding to axial-symmetry, thereby ensuring compatibility of the description with the first law. Our results reveal that the reference frame of a given thermodynamic representation is directly encoded in the Killing vector generating the horizon.
In contrast, mass and volume are treated as potential quantities, defined only up to a gauge. This freedom reflects the role of the cosmological constant as an external field. A consistent first law emerges once the gauge is properly fixed.

This perspective clarifies, on physical grounds, why distinct formulations of KadS thermodynamics assign different values to temperature, angular velocity, and volume. Specifically, it provides a systematic interpretation of isohomogeneous transformations: they correspond to changes of reference frame, which fix the kinematic terms, combined with gauge transformations, that affect the dynamical ones. 

We also demonstrated that the quantum statistical relation restricts the set of homogeneous KadS descriptions. In particular, those satisfying this condition reduce to the usual Kerr thermodynamics in the limit of vanishing cosmological constant. This ensures the consistency with the asymptotically flat case, or to the traditional Schwarzschild-adS thermodynamics in the non-rotating limit. Moreover, the UTT  emerges as the unique thermodynamic representation tied to the frame co-rotating with infinity. Similarly, the ATT is uniquely selected when imposing that the thermodynamic volume must coincide with the geometric one, defined as a vector volume.

It is instructive to place our results in the context of the relevant literature. An adequate thermodynamic treatment can be derived from the original proposals presented in \cite{caldarelli2000thermodynamics,gibbons2005first}. However, a notable issue with this formulation is its failure to reproduce an adequate Smarr formula and incorporate scale invariance. To address this, it has been proposed to interpret the cosmological constant as a thermodynamic variable \cite{kastor2009enthalpy,kubizvnak2012p, kubizvnak2017black,dolan2011compressibility,dolan2011pressure,dolan2012pdv, xiao2024extended, cai2005thermodynamics, campos2025black, gao2023general}. Nevertheless, these works rely on scaling arguments in which Komar integrals are identified with the Euler relation for homogeneous functions. As we have previously demonstrated \cite{campos2024generating}, such an identification cannot always be made. These unresolved issues keep the thermodynamics of Kerr-adS spacetimes an active area of research.
Altogether, our results provide a coherent physical interpretation for the multiplicity of KadS thermodynamic descriptions, grounding them in geometric and quantum-statistical considerations. They suggest that black-hole thermodynamics in adS might be understood as a combination of frame dependence (kinematics) and gauge fixing (dynamics).

The presented developments open new avenues for further research. For instance, since imposing the validity of the QSR indicates an agreement between the entropy derived from the Euclidean formalism and the generalized Iyer-Wald entropy (No{\"e}ther-Wald charge) \cite{wald1993black}, a natural extension of this work is to investigate whether this consistency holds in broader contexts. Other areas for further exploration include more complex asymptotically adS black holes, generalizations with charges, and higher-derivative gravity theories, with potential applications in holography and semiclassical gravity.

\appendix

\section{Characterization of generating functions that preserve the QSR}

\label{app:functionG}

In this appendix we demonstrate that the generating function $g$ associated with an isohomogeneous transformation takes the form
\begin{equation}
g = G \left( \mathcal{F}\right) ~, ~~ \textrm{with} ~~ \mathcal{F}=\frac{J^{2}P}{M_{0}^{2}}~.
\label{h-sol}
\end{equation}

Consider an isohomogeneous transformation, not necessarily exact, that relates the potential $M_{1}$ to the exact potential $M_{0}$, $M_{1}=gM_{0}$. Furthermore, consider that both potentials respect QSR, that is, $g$ is a transformation that preserves QSR. Thus, $g$ obeys Eq.~\eqref{condition qsr}, with $I$ given by Eq.~\eqref{eq I}.
Using the expression~\eqref{eq I} for $I$, we can multiply Eq.~\eqref{condition qsr} by $T_{0}$ and write
\begin{equation}
\frac{\partial g}{\partial S}\left( \frac{M_{0}}{2}-P\frac{\partial M_{0}}{%
\partial P}\right) +\frac{\partial M_{0}}{\partial S}\frac{\partial g}{%
\partial P}P=0~.  
\label{2}
\end{equation}
Furthermore, due to the homogeneity of $M_{0}$ and $g$ [Eqs.~\eqref{original smarr} and \eqref{homoeneity of g} respectively], we have
\begin{eqnarray}
&&\frac{M_{0}}{2}=S\frac{\partial M_{0}}{\partial S}+J\frac{\partial M_{0}}{%
\partial J}-P\frac{\partial M_{0}}{\partial P}~,  \label{3} \\
&&S\frac{\partial g}{\partial S}+J\frac{\partial g}{\partial J}-P\frac{%
\partial g}{\partial P}=0~.  \label{1}
\end{eqnarray}

Let $\omega \equiv \mathrm{d}g$ 
be the 1-form obtained from the exterior derivative of the function 
$g=g\left( S,J,P\right) $. We introduce the vector fields $L_1$ and $L_2$, that can act as operators in functions on the manifold,
\begin{equation}
L_{1} = 
S \frac{\partial}{\partial S}  
+ J \frac{\partial}{\partial J} 
- P \frac{\partial}{\partial P} ~,\qquad 
L_{2} = 
\left( \frac{M_{0}}{2}-P\frac{\partial M_{0}}{\partial P}\right) 
\frac{\partial}{\partial S} 
+ P \frac{\partial M_{0}}{\partial S} 
\frac{\partial}{\partial P } ~ .
\end{equation}
Equations~\eqref{2} and \eqref{1} can be written as
\begin{eqnarray}
\omega \left( L_{1}\right)  &=&\omega \left( L_{2}\right) =0~,  \label{k} \\
L_{1}\left( g\right)  &=&L_{2}\left( g\right) =0~.  \label{L}
\end{eqnarray}
That is, the vectors $L_{1}$ and $L_{2}$ belong to the kernel of $\omega$, while $g$ itself lies in the kernel of both $L_{1}$ and $L_{2}$. 
By Cartan's fundamental identity, it follows that
\begin{equation}
L_{1}\left( \omega \left( L_{2}\right) \right) -L_{2}\left( \omega \left(
L_{1}\right) \right) -\omega \left( \left[ L_{1},L_{2}\right] \right)
=-\omega \left( \left[ L_{1},L_{2}\right] \right) =\mathrm{d}\omega =0~.
\label{dw=0}
\end{equation}
In the first step of Eq.~\eqref{dw=0} we use Eq.~\eqref{k}, and in the second step, we use the fact that $\omega$ is exact.

The next step is to verify that vector fields $L_{1}$ and $L_{2}$ are linearly independent. For this purpose, let us determine the rank of the matrix $m$, where
\begin{equation}
m=\left( 
\begin{array}{ccc}
S & J & -P \\ 
\frac{M_{0}}{2}-P\frac{\partial M_{0}}{\partial P} & 0 & P\frac{\partial
M_{0}}{\partial S}
\end{array}
\right) ~.
\end{equation}
The minors $m_{12}$ and $m_{23}$ of the matrix $m$ are
\begin{eqnarray}
m_{12} &=&\det \left( 
\begin{array}{cc}
S & J \\ 
\frac{M_{0}}{2}-P\frac{\partial M_{0}}{\partial P} & 0
\end{array}
\right) =-J\left( \frac{M_{0}}{2}-P\frac{\partial M_{0}}{\partial P}\right)
~, 
\label{4-1}
\\
 m_{23} &=&\det \left( 
\begin{array}{cc}
J & -P \\ 
0 & P\frac{\partial M_{0}}{\partial S}
\end{array}
\right) =JP\frac{\partial M_{0}}{\partial S}~.  
\label{5}
\end{eqnarray}
We are assuming asymptotically anti-de Sitter rotating black holes, so that $J \ne 0$, $P \ne 0$ and $M_{0} \ne 0$. Moreover, the Smarr relation~\eqref{original smarr} implies that $\partial M_{0} / \partial P \ne 0 $ and $\partial M_{0} / \partial S \ne 0 $. From Eqs.~\eqref{4-1} and \eqref{5}, we have that $m_{12} \ne 0$ and $m_{23} \ne 0$, hence the rank of $m$ is 2. Consequently the vector fields $L_{1}$ and $L_{2}$ are linearly independent.

Given the linear independence of $L_{1}$ and $L_{2}$, we can invoke Frobenius's theorem to conclude that $g$ has only one free parameter. That is, the function $g$ depends only on one invariant combination of the independent variables. Direct verification using Eq.~\eqref{3} shows that this combination is given by $g = G ( \mathcal{F} )$, where $\mathcal{F}$ is indicated in Eq.~\eqref{h-sol}. This completes our demonstration.

We supplement the derivation in this appendix with comments that highlight the important role of dimensionality in KadS thermodynamics.
In the thermodynamic space with $n$ independent variables $\left\{ X_{i}\right\}$ of homogeneity $\left\{ \alpha_{i}\right\}$, there will always be two constraints (two vectors) $L_{1,2}$: one given by the homogeneity of $g$ in Eq.~\eqref{1}, and the other by the QSR in Eq.\eqref{2}. 
While $L_{1}$ necessarily takes the form
\begin{equation}
	L_{1}=\sum_{i=1}^{n} \alpha_{i}\frac{\partial }{\partial X_{i}}~,
\end{equation}
the explicit form of the vector $L_{2}$ depends on the number of dynamic (gauge-dependent) and kinematic (frame-dependent) variables. The maximum rank of the matrix formed by the coefficients of these operators is $2$, and if the equations remain involutive (that is, if the theory in question is integrable) the minimum number of independent parameters will be $(n-2)$. For $n=2$, as we saw  with the Schwarzschild-adS and Kerr spacetimes, there will be zero free parameters, implying a unique thermodynamic description. The number of free parameters can be reduced further by imposing additional constraints. For example, for $n=3$, any other independent restriction on the derivatives of $g$ (e.g., fixing $T_{1}-T_{0}$, $V_{1}-V_{0}$, $\Omega _{1}-\Omega _{0}$) generates an additional constraint, reducing the number of free parameters to zero. This is the general idea behind the conclusions following Eqs.~\eqref{sads and kerr unique} and~\eqref{qsr T omega V 2}, concerning the unique status of the Schwarzschild-adS and Kerr thermodynamics and the role of the factor $\mathcal{F}$.

\section{Potential volume and other charges}

\label{app: forms}

This appendix reviews elements of the formalism involving potential volume and other charges that are relevant to the present work.
A very general notion of charge can be associated with the flow of divergent free vector fields. In other words, the charge in a region is equal to the field flux across the boundary of the region. The local flux of a field $J=J^{\mu }\partial _{\mu }$ across a infinitesimal volume of a four dimensions (metric) manifold (development can be extended to higher dimensions) can be written as
\begin{equation}
\imath _{J}\left( \star 1\right) ~,\ \star 1=\mathrm{vol}=\sqrt{\left\vert
g\right\vert }\mathrm{d}x^{0}\wedge \mathrm{d}x^{1}\wedge \mathrm{d}x^{2}\wedge \mathrm{d}x^{3}~,
\end{equation}%
where $\imath _{J}\left( \star 1\right) =J\cdot \left( \star 1\right) $ is
the interior product of field $J$ with the $4$-form $\star 1$. In addition,
for any 1-form $j=J_{\mu }\mathrm{d}x^{\mu }$ we have%
\begin{equation}
\star \mathrm{d}\star j=\nabla^{\mu }J_{\mu }\Rightarrow \mathrm{d}\star j=-\nabla^{\mu
}J_{\mu }\star 1~.  \label{m6}
\end{equation}
From the Cartan formula, we see that 
\begin{equation}
\mathcal{L}_{J}\left( \star 1\right) =\mathrm{d}\left( \imath _{J}\left( \star
1\right) \right) +\imath _{J}\mathrm{d}\left( \star 1\right) =\mathrm{d}
\left( \imath _{J}\left( \star 1\right) \right) =\nabla_{\mu }J^{\mu }\left(
\star 1\right) ~,  \label{m0}
\end{equation}%
where $\mathcal{L}$ and $\mathrm{d}$ are the Lee and exterior derivatives, respectively. From the above expressions we have%
\begin{equation}
\mathrm{d}\left( \imath _{J}\left( \star 1\right) \right) =\nabla_{\mu
}J^{\mu }\left( \star 1\right) = -\mathrm{d}\star j~,
\end{equation}
where $j$ and $J$ are related by the musical isomorphism, $J=j^{\#}~,\
j=J^{\flat }$. In local coordinates
\begin{align}
\begin{split}
&  \star j=\frac{\sqrt{\left\vert g\right\vert }}{3!}J_{\mu }\varepsilon
_{\;\alpha \beta \gamma }^{\mu }\mathrm{d}x^{\alpha }\wedge \mathrm{d}x^{\beta }\wedge
\mathrm{d}x^{\gamma }=J^{\mu }\mathrm{d}\Sigma _{\mu } ~,
\\
& \mathrm{d} \Sigma _{\mu }\equiv \frac{\sqrt{\left\vert g\right\vert }}{3!}
\varepsilon _{\mu \nu \alpha \beta }\mathrm{d}x^{\nu }\wedge \mathrm{d}x^{\alpha }\wedge
\mathrm{d}x^{\mu } \,.
\end{split}
\label{quantidades BL1}
\end{align}

We now particularize the case of a divergent free field $J$, $\nabla_{\mu
}J^{\mu }=0$, i.e, when $J$ represents a conserved current. In this case,
from Eq.~\eqref{m6}, we have
\begin{equation}
\nabla_{\mu }J^{\mu }=0\Longleftrightarrow \mathrm{d}\star j=0~,
\end{equation}%
and the Poincar\'{e} lemma implies that, locally,
\begin{equation}
\mathrm{d}\star j=0\Rightarrow \star j=\mathrm{d}\omega =J^{\mu }\mathrm{d}\Sigma
_{\mu }~.  \label{m3}
\end{equation}%
As a result,\footnote{%
We are suppressing the pullback notation in the integrals.}
\begin{equation}
\int_{\Sigma }J^{\mu }\mathrm{d}\Sigma _{\mu }=\int_{\Sigma }\imath _{J}\left( \star
1\right) =\int_{\Sigma }\mathrm{d}\omega =\int_{\partial \Sigma }\omega ~.
\label{3.24}
\end{equation}
For any contractible surface $\Sigma$ in $\mathbb{R}^{3}$, we can define a charge in region $\Sigma$, associated with conserved current $J$, as
\begin{equation}
Q_{J}\equiv \int_{\Sigma }J^{\mu }\mathrm{d}\Sigma _{\mu }=\int_{\partial \Sigma
}\omega ~.  \label{carga}
\end{equation}

A common case is when $J$ is associated with a Killing vector $K=K^{\mu }\partial _{\mu }$,
\begin{equation}
2\nabla_{\mu }\nabla_{\nu }K^{\mu }=2R_{\mu \nu }K^{\mu }=J_{\nu }~.
\label{2.19}
\end{equation}
In this case,
\begin{equation}
\star \mathrm{d}\star \mathrm{d}k=j\Rightarrow \mathrm{d}\star \mathrm{d}
k=\star ^{-1}j~,~k=K^{\flat }~.  
\label{2.20}
\end{equation}
Therefore, $J=j^{\#}$ is a conserved current,
\begin{equation}
\star ^{-1}j=\mathrm{d}\star \mathrm{d}k\Rightarrow \mathrm{d}\star
j=0\Rightarrow \nabla_{\mu }J^{\mu }=0~.
\end{equation}
In this case, $\omega =\star \mathrm{d}k$ in Eq.~\eqref{carga} and the expression for $Q_{J=K}$\ is known as a Komar integral. Note the similarity of $\mathrm{d}\star \mathrm{d}k=\star ^{-1}j$ with the non-homogeneous Maxwell equation by identify $k$ with the electromagnetic vector potential. In fact, the tensor 
\begin{equation}
F_{\mu \nu }=\nabla_{\mu }k_{\nu }-\nabla_{\nu }k_{\mu }=2\nabla_{\mu
}k_{\nu }~,
\end{equation}%
is known as the Ernst $2$-form $F=\mathrm{d}k$, and the first Bianchi
identity is given by the homogeneous equation $\mathrm{d}F=0$.

The fact that, for any closed $2$-form $\eta $ (that is, $\mathrm{d}\eta =0$), 
\begin{equation}
\omega ^{\prime }=\omega +\eta \Rightarrow \mathrm{d}\omega ^{\prime }= 
\mathrm{d}\omega =\imath _{J}\left( \star 1\right) ~,
\end{equation}
represents the gauge freedom of the theory. Since $\Sigma$ in Eq.~\eqref{carga} is a contractible region, the gauge $\eta $ does not modify the charge
\begin{equation}
\int_{\partial \Sigma }\eta =0~.
\end{equation}
However, issues can arise when the boundary of $\Sigma $ can be split into two regions $A$ and $B$ ($A\cup B=\partial \Sigma $). For example, in the case of black holes, two concentric spheres around the singularity. In this case, 
\begin{equation}
\int_{\partial \Sigma =A\cup B}\eta =\int_{A}\eta +\int_{B}\eta =0\text{
with }\ \int_{A}\eta =-\int_{B}\eta \neq 0~.  \label{m8}
\end{equation}%
The quantity 
\begin{equation}
q_{J}=\int_{A}\omega ^{\prime }=\int_{A}\left( \omega +\eta \right) ~,
\end{equation}
may be gauge dependent, even if region $A$ is the boundary of  region ($A=\partial C$), as the cited concentric sphere around the singularity. This happens if $A$ is non-contractile region, and the cohomology group of the manifold is non-trivial (i.e., if there are closed forms which are not exact).

In the present work, one case of interest is when $J$ is a Killing field $K$%
, $J=K$. Therefore, $K$ is a conserved current $\nabla_{\mu }K^{\mu }=0$.
For the above-mentioned case where $\partial \Sigma =A\cup B$, $A=\partial C$%
, $A$ and $B$ are the concentric spheres around the singularity of the black
hole, the quantity 
\begin{equation}
q_{K}=\int_{\partial C}\omega ~,
\end{equation}%
is the potential volume and the improper integral 
\begin{equation}
V_{K}=\lim_{B\rightarrow 0}\int_{\Sigma }K^{\mu }\mathrm{d}\Sigma _{\mu }~,
\end{equation}%
is the vector volume.

\begin{acknowledgments}
	
T.~L.~C. acknowledges the support of Coordena\c{c}\~ao de Aperfei\c{c}oamento de Pessoal de N\'{\i}vel Superior (CAPES) -- Brazil, Finance Code 001.
C.~M. acknowledges the support of S\~ao Paulo Research Foundation (FAPESP) -- Brazil, Grant No.~2022/07534-0.
	
\end{acknowledgments}


\begin{thebibliography}{99}
	
	
	%22
	\bibitem{Maldacena:1997re} J.~M.~Maldacena, The Large N limit of
	superconformal field theories and supergravity, Adv. Theor. Math. Phys. 
	\textbf{2}, 231 (1998).  arXiv:hep-th/9711200
	
	%23
	\bibitem{Witten:1998qj} E.~Witten, Anti-de Sitter space and holography, Adv. Theor. Math. Phys. \textbf{2}, 253 (1998).  arXiv:hep-th/9802150
	
	%24
	\bibitem{Amado:2017kao} J.~B.~Amado, B.~C.~da Cunha, and E.~Pallante, On the
	KadS/CFT correspondence, J. High Energy Phys. \textbf{08}, 094 (2017).
	arXiv:1702.01016
	
	%25
	\bibitem{Guica:2008mu} M.~Guica, T.~Hartman, W.~Song, and A.~Strominger, The Kerr/CFT correspondence, Phys. Rev. D \textbf{80}, 124008 (2009).  
	arXiv:0809.4266
	
	%26
	\bibitem{Brown:1994gs} J.~D.~Brown, J.~Creighton, and R.~B.~Mann,
	Temperature, energy and, heat capacity of asymptotically anti-de Sitter
	black holes, Phys. Rev. D \textbf{50}, 6394 (1994).  
	arXiv:gr-qc/9405007
	

	%31
	\bibitem{Cardoso:2013pza} V.~Cardoso, \'{O}.~J.~C.~Dias, G.~S.~Hartnett,
	L.~Lehner, and J.~E.~Santos, Holographic thermalization, quasinormal modes
	and superradiance in KadS, J. High Energy Phys. \textbf{04}, 183 (2014).
	arXiv:1312.5323
	
	%35
	\bibitem{bardeen1973four} J.~M.~Bardeen, B.~Carter, and S.~W.~Hawking, The
	Four laws of black hole mechanics, Commun. Math. Phys. \textbf{31}, 161
	(1973).
		
	%37
	\bibitem{gauntlett1999black} J.~P.~Gauntlett, R.~C.~Myers, and
	P.~K.~Townsend, Black holes of D=5 supergravity, Class. Quant. Grav. \textbf{16}, 1 (1999). 
	arXiv:hep-th/9810204
	
	
	
	%38
	\bibitem{townsend2001first} P.~K.~Townsend, and M.~Zamaklar, The First law of black brane mechanics, Class. Quant. Grav. \textbf{18}, 5269 (2001). 
	arXiv:hep-th/0107228
	
	%21
	\bibitem{Gubser:1998bc} S.~S.~Gubser, I.~R.~Klebanov, and A.~M.~Polyakov,
	Gauge theory correlators from non-critical string theory, Phys. Lett. B 
	\textbf{428}, 105 (1998).  arXiv:hep-th/9802109
	
	
	
	%27
	\bibitem{Elias:2018yct} W.~S.~Elias, C.~Molina, and M.~C.~Baldiotti.
	Thermodynamics of bosonic systems in anti-de Sitter spacetime, Phys. Rev. D \textbf{99} 084028 (2019).  
	arXiv:1803.05921
	
	%28
	\bibitem{Hemming:2007yq} S.~Hemming, and L.~Thorlacius, Thermodynamics of
	Large AdS Black Holes, J. High Energy Phys. \textbf{11}, 086 (2007).
	arXiv:0709.3738
	
	%30
	\bibitem{Louko:1996dw} J.~Louko, and S.~N.~Winters-Hilt, Hamiltonian
	thermodynamics of the Reissner-Nordstrom anti-de Sitter black hole, Phys.
	Rev. D \textbf{54}, 2647 (1996).  
	arXiv:gr-qc/9602003
	
	%33
	\bibitem{Fontana:2018drk} W.~B.~Fontana, M.~C.~Baldiotti, R.~Fresneda, and C.~Molina, Extended quasilocal thermodynamics of Schwarzschild-anti de Sitter black holes, Annals Phys. \textbf{411}, 167954 (2019).   arXiv:1806.05699
	
	%34
	\bibitem{Baldiotti:2017ywq} M.~C.~Baldiotti, R.~Fresneda, and C.~Molina, A
	Hamiltonian approach for the thermodynamics of AdS black holes, Annals Phys. 
	\textbf{382}, 22 (2017). 
	arXiv:1701.01119
	
	%40
	\bibitem{chrusciel2015hamiltonian} P.~T.~Chru{\'s}ciel, J.~Jezierski, and
	J.~Kijowski, Hamiltonian dynamics in the space of asymptotically Kerr-de
	Sitter spacetimes, Phys. Rev. D \textbf{92}, 084030 (2015).  arXiv:1507.03868
	
	%20
	\bibitem{dolan2011pressure} B.~P.~Dolan, Pressure and volume in the first
	law of black hole thermodynamics, Class. Quant. Grav. \textbf{28}, 235017
	(2011).  arXiv:1106.6260
	
	
	
	
	
	
	%32
	\bibitem{Lemos:2015zma} J.~P.~S.~Lemos, F.~J.~Lopes, M.~Minamitsuji, and
	J.~V.~Rocha, Thermodynamics of rotating thin shells in the BTZ spacetime,
	Phys. Rev. D \textbf{92}, 064012 (2015).  arXiv:1508.03642
	
	
	
	%1
	\bibitem{caldarelli2000thermodynamics} M.~M.~Caldarelli, G.~Cognola, and
	D.~Klemm, Thermodynamics of Kerr-Newmann-AdS black holes and conformal field
	theories, Class. Quant. Grav. \textbf{17}, 399 (2000).  
	arXiv:hep-th/9908022
	
	%2
	\bibitem{gibbons2005first} G.~W.~Gibbons, M.~J.~Perry, and C.~N.~Pope, The
	first law of thermodynamics for Kerr-anti de Sitter black holes, Class.
	Quant. Grav. \textbf{22}, 1503 (2005).  arXiv:hep-th/0408217
	
	%3
	\bibitem{cvetivc2011black} M.~Cveti{\v c}, G.~W.~Gibbons, D.~Kubiz{\v n}{\'a}k, and C.~N.~Pope, Black hole enthalpy and an entropy inequality for the thermodynamic volume, Phys. Rev. D \textbf{84}, 024037 (2011).   arXiv:1012.2888
	
	%4
	\bibitem{kubizvnak2017black} D.~Kubiz{\v{n}}{\'a}k, R.~B.~Mann, and M.~Teo, Black hole chemistry: Thermodynamics with Lambda, Classical Quantum Gravity \textbf{34}, 063001 (2017). 
	arXiv:1608.06147
	
	%5
	\bibitem{kubizvnak2012p} D.~Kubiz{\v{n}}{\'a}k, R.~B.~Mann, and B.~Robert,
	P-V criticality of charged AdS black holes, J. High Energy Phys. \textbf{07}, 033 (2012). 
	arXiv:1205.0559
	
	%6
	\bibitem{dolan2011compressibility} B.~P.~Dolan, Compressibility of rotating black holes, Phys. Rev. D \textbf{84}, 127503 (2011).  arXiv:1109.0198
	
	%7
	\bibitem{xiao2024extended} Y.~Xiao, Y.~Tian, and Y.~Liu, Extended black hole thermodynamics from extended Iyer-Wald formalism,  Phys. Rev. Lett. \textbf{132}, 021401 (2024). 
	arXiv:2308.12630
	
	%8
	\bibitem{dolan2012pdv}
	B.~P.~Dolan, Where is the PdV in the first law of black hole thermodynamics? \textit{Open Questions in Cosmology} (IntechOpen, 2012). arXiv:1209.1272
	
	%9
	\bibitem{cai2005thermodynamics} R.-G. Cai, L.-M. Cao, and D.-W. Pang, Thermodynamics of Dual CFTs for Kerr-AdS Black Holes, Phys. Rev. D \textbf{72} 044009 (2005). arXiv:hep-th/0505133
	
	%10
	\bibitem{hawking1999rotation} S.~W.~Hawking, C.~J.~Hunter, and
	M.~M.~Taylor-Robinson, Rotation and the AdS/CFT correspondence, Phys. Rev. D \textbf{59}, 064005 (1999).  arXiv:hep-th/9811056
	
	%11
	\bibitem{gao2023general} Y.~Gao, Z.~Di, and S.~Gao, General mass formulas for charged KadS black holes, 
	%
	Phys. Scr. \textbf{99}, 095022 (2024).
	arXiv:2304.10290
	
	%12
	\bibitem{iyer1994some} V.~Iyer, and R.~M.~Wald, Some properties of the
	Noether charge and a proposal for dynamical black hole entropy, Phys. Rev. D \textbf{50}, 849 (1994). 
	arXiv:gr-qc/9403028
	
	%13
	\bibitem{campos2025black} T.~L.~Campos, M.~C.~Baldiotti, and C.~Molina, Black-hole thermodynamics from gauge freedom in extended Iyer-Wald formalism,  Universe \textbf{11}, 2015 (2025).  arXiv:2507.03751
	
	%14
	\bibitem{ballik2013vector} W.~Ballik, and K.~Lake, Vector volume and black holes, Phys. Rev. D, \textbf{88}, 104038 (2013).  arXiv:1310.1935
	
	%15
	\bibitem{campos2024generating} T.~L.~Campos, M.~C.~Baldiotti, C.~Molina, Generating Kerr--anti--de Sitter thermodynamics, Phys. Rev. D \textbf{110}, 024049 (2024). 
	arXiv:2407.09610
	
	%16
	\bibitem{kastor2009enthalpy} D.~Kastor, S.~Ray, and J.~Traschen, Enthalpy
	and the mechanics of adS black holes, Class. Quant. Grav. \textbf{26},
	195011 (2009).  
	arXiv:0904.2765
	
	%17
	\bibitem{gibbons1977action} G.~W.~Gibbons, and S.~W.~Hawking, Action
	integrals and partition functions in quantum gravity, Phys. Rev. D \textbf{15} 2752 (1977).
	
	%18
	\bibitem{gibbons1978black} G.~W.~Gibbons, M.~J.~Perry, Black holes and
	thermal Green functions, Proceedings of the Royal Society of London. A.
	Mathematical and Physical Sciences \textbf{358}, 467 (1978).
	
	%19
	\bibitem{griffiths2009exact} J.~B.~Griffiths and J.~Podolsk{\`y}, \textit{Exact space-times in Einstein's general relativity} (Cambridge University Press, 2009).
	
	
	
	%39
	\bibitem{santiago2018tolman} J.~Santiago, and M.~Visser, Tolman-like
	temperature gradients in stationary spacetimes, Phys. Rev. D \textbf{98}
	064001 (2018).  
	arXiv:1807.02915
	
	
	
	%20b
	\bibitem{wald1993black} R.~M.~Wald, Black hole entropy is the Noether charge,  Physical Review D \textbf{48}, R3427 (1993).  arXiv:gr-qc/9307038
	
	
	
	
	
	
	
	
	
	
\end{thebibliography}
\end{document}